\documentstyle[12pt,fleqn]{article} %
\def\stot{\sigma_{\rm tot}}
\def\pbar {\bar p}

\def\ie{{\it i.e.,\ }}

\begin{document}
\input FEYNMAN
\renewcommand\thepage{\ }
\newcommand{\BS}{\bigskip}
\newcommand{\eq}[1]{eq.~(\ref {#1})}
\newcommand{\Em}[1]{{\em {#1}}}
\def\beq{\begin{equation}}
\def\eeq{\end{equation}}
\def\beqa{\begin{eqnarray}}
\def\eeqa{\end{eqnarray}}
\def\rd{{\mathrm d}}
\def\thbar{{\bar\theta}}
\def\intd4x{\int{\rd}^4x}
\def\lcal{{\cal L}}
\def\hc{\mathrm{h.c.}}
\def\STr{\mathrm{STr}}
\def\Tr{{\mbox {\rm Tr}}\,}
\def\dmuu{\partial^\mu}
\def\dmud{\partial_\mu}
\def\da{{\dot{a}}}
\def\db{{\dot{b}}}
\def\m32{{m_{3/2}}}
\newcommand{\ket}[1]{|#1\rangle}
\newcommand{\bra}[1]{\langle#1|}
\newcommand{\vev}[1]{\left\langle#1\right\rangle}
\newcommand{\eqr}[1]{~(\ref{eq:#1})}
\newcommand{\Sl}[1]{/ \mskip-10mu {#1}} %

\begin{titlepage} %
\newcommand\reportnumber{489} %
\newcommand\mydate{September, 1995} %
\newlength{\nulogo} %
\settowidth{\nulogo}{\small\sf{N.U.H.E.P. Report No. \reportnumber}}
\title{{\ }\vspace{-1.in}\\
\hfill\fbox{{\parbox{\nulogo}{\small\sf{Northwestern University: \\
N.U.H.E.P. Report No. \reportnumber\\ \mydate
}}}}\vspace{1in} \\
{Are We Really Measuring the $\rho$-Value?}}
\author{M.~M.~Block
\thanks{Work partially supported by Department of Energy grant
DA-AC02-76-Er02289 Task B.} \\
{\small\em Department of Physics and Astronomy,} \vspace{-5pt} \\ %
{\small\em Northwestern University, Evanston, IL 60208}\\
\vspace{.5in}\\
{\small \sf Paper presented by Martin~M.~Block}\\[-4pt]
{\small \sf at the}\\[-4pt]
{\small \sf VIth Blois Workshop, Chateau de Blois, France,  June,
1995}\\[-4pt]
}
\vfill
\date{} %
\maketitle
\begin{abstract} %
\noindent The ``normal'' analysis of $\pbar p$ and $pp$ elastic scattering
uses
a `spinless'
Coulomb amplitude, \ie a Rutherford amplitude ($2\sqrt{\pi}\alpha/t$)
multiplied by a Coulomb
form factor $G^2(t)$, an {\em ansatz} that pretends that the nucleon does not
have any magnetic
scattering.  In this note, we investigate the role of the anomalous magnetic
moment of the
nucleon, $\kappa \approx 1.79$.  Given the method of analysis currently used
by
most published
experiments, we conclude that the current experimentally inferred values  of
$\rho$ for $\pbar p$ should be {\em systemeatically lowered} by $\approx
0.005$---0.0100 and, correspondingly,  the $\rho$ values for $pp$ should be
{\em systematically raised} by the same amount. We discuss the theoretical
uncertainties and
a method of experimentally minimizing them.
\end{abstract}  %
\end{titlepage} %
\pagenumbering{arabic}
\renewcommand{\thepage}{-- \arabic{page}\ --}  %
\renewcommand{\thesection}{\Roman{section}}  %
\renewcommand{\thesubsection}{\Roman{section}.\alph{subsection}}
\section{Introduction}
We will only calculate electromagnetic amplitudes accurate to order $\alpha$,
\ie one-photon exchange diagrams. Further, we will consider only high energy
scattering ($E_{\rm lab} \gg m$, where  $m$ is the nucleon mass) in the region
of small $|t|$,
where $t$ is the squared 4-momentum transfer. We will measure $m$ and  $E_{\rm
lab}$ in GeV and
$t$ in (GeV)$^2$, and will use $\hbar=c=1$.
\subsection{`Spinless' Coulomb Scattering}
If we consider `spinless' proton-antiproton Coulomb scattering, the relevant
Feynman diagram
is shown in Fig. \ref{spinless}.
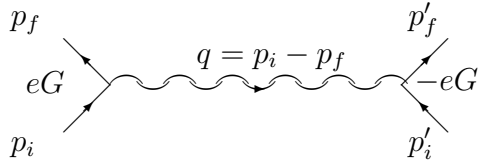
\begin{figure}[hbt]
\begin{picture}(30000,10000)
\THINLINES
\drawline\fermion[\NW\REG](15000,5000)[2500]
\drawarrow[\NW\ATBASE](\pmidx,\pmidy)%
\advance\pbacky by 500
\advance\pbackx by -2000
\put(\pbackx,\pbacky){$p_f$}%
\drawline\fermion[\SW\REG](15000,5000)[2500]
\advance\pbackx by -2000
\put(\pbackx,2500){$p_i$}%
\drawarrow[\NE\ATBASE](\pmidx,\pmidy)%
\drawline\photon[\E\REG](15000,5000)[11]
\advance\pfrontx by -3200 %
\advance\pfronty by -300
\put(\pfrontx,\pfronty){$eG$}
\advance\pbackx by 500
\advance\pbacky by -200
\put(\pbackx,\pbacky){$-eG$}
\advance\pmidy by -250
\advance\pmidy by 100
\drawarrow[\E\ATBASE](\pmidx,\pmidy)%
\advance\pmidx by -2250
\advance\pmidy by 1000
\put(\pmidx,\pmidy){$q=p_i-p_f$}
\drawline\fermion[\NE\REG](\photonbackx,\photonbacky)[2500]
\advance\pbacky by 500
\advance\pbackx by -1500
\put(\pbackx,\pbacky){$p'_f$}
\drawarrow[\NE\ATBASE](\pmidx,\pmidy)%
\drawline\fermion[\SE\REG](\photonbackx,\photonbacky)[2500]
\advance\pbackx by -1500
\put(\pbackx,2500){$p'_{i}$}%
\drawarrow[\NW\ATBASE](\pmidx,\pmidy)%
\end{picture}
\caption{\protect{\footnotesize {One-Photon Feynman Diagram for `spinless'
Proton---Anti-Proton
Coulomb Scattering,
$p_i +p'_i\rightarrow p_f+p'_f$, with Couplings
$eG$ and $-eG$ , where $G(q^2)$ is the
Electromagnetic Charge
Form Factor of the Nucleon.}}}\label{fig:spinless}
\end{figure}
The electromagnetic differential cross section is readily shown to be
\begin{eqnarray}
\frac{d\sigma}{dt}&=&
4\pi G^4(t)
\frac{\alpha^2}{\beta_{\rm lab}^2t^2}\times
\left(1-\frac{|t|}{4mE_{\rm lab}}\right)^2\nonumber\\
&=&\pi
\left|
\frac{\mp 2G^2(t)\alpha}{\beta_{\rm lab}|t|}\times
\left(1-\frac{|t|}{4mE_{\rm lab}}\right)\right|^2,
\label{spin0answer}
\end{eqnarray}
where the upper (lower) sign is for like (unlike) charges, $t$ is the
(negative) 4-momentum
transfer squared, and $m$ is the nucleon mass.

For small angle scattering, the  term
$\left(1-\frac{|t|}{4mE_{\rm lab}}\right)^2
\approx 1-\frac{|t|}{2mE_{\rm lab}}$ and
\beq
\frac{d\sigma}{dt}\approx
4\pi G^4(t)
\frac{\alpha^2}{\beta_{\rm lab}^2t^2}\times
\left(1-\frac{|t|}{2mE_{\rm lab}}\right)\label{spinless}.
\eeq
At high energies, the correction term
$\frac{|t|}{2mE_{\rm lab}}$ becomes  negligible and $\beta_{\rm lab}
\rightarrow 1$, so
\eq{spinless}  goes over into the well-known Rutherford
scattering formula,
\beq
\frac{d\sigma}{dt}
=\pi\left|
\frac{\mp 2\alpha G^2(t)}
{|t|}\right|^2
,\label{rutherford}
\eeq
where the electromagnetic charge form factor $G(t)$ is commonly parameterized
by the dipole form
\beq
G(t)=\frac{1}{\left(1-\frac{t}{\Lambda^2}\right)^2},\label{G}
\eeq
where $\Lambda^2=0.71$, if $t$ is measured in (GeV)$^2$.
We note that this is the Coulomb amplitude that is normally used in the
analysis of
$\pbar p$ and $pp$ elastic scattering, \ie  the `spinless' analysis\cite{bc}.
\subsection{$\pbar p$ Scattering, Including Magnetic Scattering}
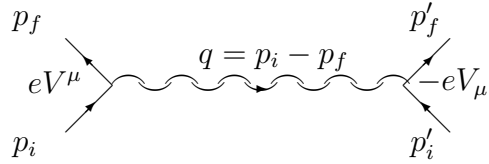
\begin{figure}[htb]
\begin{picture}(30000,10000)
\THINLINES
\drawline\fermion[\NW\REG](15000,5000)[2500]
\drawarrow[\NW\ATBASE](\pmidx,\pmidy)%
\advance\pbacky by 500
\advance\pbackx by -2000
\put(\pbackx,\pbacky){$p_f$}%
\drawline\fermion[\SW\REG](15000,5000)[2500]
\advance\pbackx by -2000
\put(\pbackx,2500){$p_i$}%
\drawarrow[\NE\ATBASE](\pmidx,\pmidy)%
\drawline\photon[\E\REG](15000,5000)[11]
\advance\pfrontx by -3200 %
\advance\pfronty by -300
\put(\pfrontx,\pfronty){$eV^\mu$}
\advance\pbackx by 500
\advance\pbacky by -200
\put(\pbackx,\pbacky){$-eV_\mu$}
\advance\pmidy by -250
\advance\pmidy by 100
\drawarrow[\E\ATBASE](\pmidx,\pmidy)%
\advance\pmidx by -2250
\advance\pmidy by 1000
\put(\pmidx,\pmidy){$q=p_i-p_f$}
\drawline\fermion[\NE\REG](\photonbackx,\photonbacky)[2500]
\advance\pbacky by 500
\advance\pbackx by -1500
\put(\pbackx,\pbacky){$p'_f$}
\drawarrow[\NE\ATBASE](\pmidx,\pmidy)%
\drawline\fermion[\SE\REG](\photonbackx,\photonbacky)[2500]
\advance\pbackx by -1500
\put(\pbackx,2500){$p'_{i}$}%
\drawarrow[\NW\ATBASE](\pmidx,\pmidy)%
\end{picture}
\caption {\protect{\footnotesize {One-Photon Feynman Diagram for Proton---Anti-
Proton
Coulomb Scattering,
$p_i +p'_i\rightarrow p_f+p'_f$, with Couplings
$eV^\mu$ and $-eV_\mu$ , where
$V^\mu=F_1\gamma^\mu+i\frac{\kappa}{2m}F_2\sigma^{\nu \mu}q_\nu$,
with form factors
$F_1(q^2)$ and $F_2(q^2)$.}}}\label{withkappa}
\end{figure}

The relevant Feynman diagram is shown in Fig. \ref{withkappa}, where magnetic
scattering
is explicitly taken into account via the anomalous
magnetic moment  $\kappa$ ($ \approx 1.79$).
The fundamental electromagnetic interaction is
\beq
eV^\mu=e\left(F_1\gamma^\mu+i\frac{\kappa}{2m}F_2\sigma^{\nu
\mu}q_\nu\right),\qquad q=p_f-p_i
\label{interaction}
\eeq
which has two form factors
$F_1(q^2)$ and $F_2(q^2)$ that are normalized to 1 at $q^2=0$.  The anomalous
magnetic moment of
the nucleons is $\kappa$, and $m$ is the nucleon mass. Because of the rapid
form factor
dependence on $t$, the annihilation  diagram  for $\pbar p$ scattering (or the
exchange
diagram for $pp$ scattering) is negligible in the small $|t|$ region of
interest and has been
ignored.
The interaction of \eq{interaction} is most simply treated by using Gordon
decomposition and
can be rewritten as
\beq
eV^\mu=e\left[(F_1+\kappa F_2)\gamma_\mu -\kappa
F_2\frac{p_f+p_i}{2m}\right].\label{interaction2}
\eeq
Thus, using \eq{interaction2}, the matrix element for the scattering is
\begin{eqnarray}
M&=&e\bar u(p_f)\left[-\kappa F_2\left(\frac{p_f+p_i}{2m}\right)_\mu
+(F_1+\kappa F_2)
\gamma_\mu\right]u(p_i)\times \frac{1}{t}\times\nonumber\\
&&\mp e\bar u(p'_f)\left[-\kappa F_2\left(\frac{p'_f+p'_i}{2m'}\right)^\mu +
(F_1+\kappa F_2)
\gamma^\mu\right]u(p'_i),\label{Mkappa}
\end{eqnarray}
where the upper (lower) sign is for $\pbar p$ ($pp$) scattering.
A straightforward, albeit laborious calculation, gives a differential
scattering cross section
\begin{eqnarray}
\frac{d\sigma}{dt}
&=&4\pi \frac{\alpha^2}{\beta_{\rm lab}^2t^2}\times \nonumber\\
&&\left\{(F_1+\kappa F_2)^4\left[1+\frac{t}{2}\left(\frac{1}{mE_{\rm lab}}
+\frac{1}{E_{\rm lab}^2}\right)+\frac{t^2}{8m^2E_{\rm
lab}^2}\right]\right.\nonumber\\
&&-\,2(F_1+\kappa F_2)^2
\left[\kappa^2F_2^2\left(1+\frac{t}{4m^2}\right)+2\kappa
F_1F_2\right]\times\nonumber\\
&&\qquad\quad\left[1+\frac{t}{2}\left(\frac{1}{mE_{\rm lab}}
+\frac{1}{2E_{\rm lab}^2}\right)\right]\nonumber\\
&&+\,\left.
\left[\kappa^2F_2^2\left(1+\frac{t}{4m^2}\right)+2\kappa F_1F_2\right]^2
\left[1+\frac{t}{2mE_{\rm lab}}+\frac{t^2}{16m^2E_{\rm lab}^2}\right]\right\}
\label{sigmaF1F2}
\end{eqnarray}
We now introduce the \Em{electric} and \Em{magnetic} form factors, $G_E(t)$
and
$G_M(t)$,
defined as
\begin{eqnarray}
G_E(t)&\equiv&F_1(t)+\frac{\kappa t}{4m^2}F_2(t)%
\qquad{\rm and}\quad G_M(t)\equiv F_1(t)+\kappa F_2(t),\label{GEGM}
\end{eqnarray}
and rewrite the differential cross section of \eq{sigmaF1F2} as
\begin{eqnarray}
\frac{d\sigma}{dt}
&=&4\pi \frac{\alpha^2}{\beta_{\rm lab}^2t^2}\times\nonumber\\
&&\left\{\left(\frac{G_E^2(t)-\frac{t}{4m^2}G_M^2(t)}
{1-\frac{t}{4m^2}}
\right)^2
\left(1+\frac{t}
{2mE_{\rm lab}}\right)
+G_M^2(t)\frac{G_E^2(t)-\frac{t}{4m^2}G_M^2(t)}{1-\frac{t}{4m^2}}
\frac{t}{2E_{\rm lab}^2}\right.\nonumber\\
&&%
\quad +\,\left.\left[G_M^4(t)
+\frac{1}{2}\left(\frac{G_E^2(t)-G_M^2(t)}{1-\frac{t}{4m^2}}
\right)^2
\right]\frac{t^2}{8m^2E_{\rm lab}^2}\right\}
\label{dsigmadtpp}
\end{eqnarray}
Note in \eq{dsigmadtpp} that there is \Em{no} cross term in $G_EG_M$, but only
squares of these
form factors.  We can parameterize these new form factors with
\beq
\begin{array}{rclcl}
G_E(t)&=&G(t)&=&\displaystyle \frac{1}{\left(1-\frac{t}{\Lambda^2}\right)^2}
\qquad{\rm where } \quad \Lambda^2=0.71,\\[7mm]
G_M(t)&=&(1+\kappa)G(t)&=&\displaystyle \frac{1+\kappa}{\left(1-
\frac{t}{\Lambda^2}\right)^2},
\end{array}
\label{GEGMparameter}
\eeq
with $t$ in GeV/c$^2$, and where $G(t)$ is the dipole form factor already
defined in
\eq{G}, \ie the form factor that is traditionally used in experimental
analyses\cite{bc}.

We now expand \eq{dsigmadtpp} for \Em{very small} $|t|$, and find that
\begin{eqnarray}
\frac{d\sigma}{dt}
&\approx &4\pi \frac{\alpha^2}{\beta_{\rm lab}^2t^2}G^4(t)%
\left\{1-\kappa(\kappa +2)\frac{t}{2m^2}+\frac{t}{2mE_{\rm lab}}
+(\kappa +1)^2\frac{t}{2E_{\rm lab}^2}\right\}
,\label{dsigmadtppsmallt}
\end{eqnarray}
where the \Em{new term} in $t$, compared to \eq{spinless}, is
$-\frac{\kappa(2+\kappa)}{2m^2}t +\kappa(\kappa +1)\frac{t}{2E_{\rm lab}^2}
\approx 1+3.86|t|-\frac{3.39}{E_{\rm lab}^2}|t|$,
and is due to the anomalous magnetic moment of the proton (antiproton). To get
an estimate of
its effect, we note that $G^4(t)\approx 1-11.26|t|$, in our units where $t$ is
in GeV/c$^2$.
We note that  the new term
is \Em{not negligible} in comparison to the squared form factor, reducing
the form factor effect by about 35\%
if the
energy $E_{\rm lab}$
is large compared to $m$. In this limit, we find that independent of the
energy
$E_{\rm lab}^2$,
\beq
\frac{d\sigma}{dt}
\approx 4\pi \frac{\alpha^2}{t^2}G^4(t)\left\{
1+3.86|t|\right\}
,\label{epsmallt2}
\eeq
and is to be compared with the `spinless' Rutherford formula of
\eq{rutherford}.
\section{Effects on Experimental Analysis of Elastic Scattering}
UA4/2 has recently made a precision measurement\cite{ua42} of $\bar p$-$p$
scattering at
$\sqrt s=541$ GeV, at the S$\bar pp$S at CERN, in order to extract the
$\rho$ value for elastic scattering.  We now reanalyze this experiment, taking
into account
the magnetic scattering. They constrain the total cross section by an
independent
measurement\cite{ua4old} of $(1+\rho^2)\stot=63.3\pm 1.5$ mb. For their
published
$\rho$-value of $0.135\pm0.015$, this implies that they fix the total cross
section  at
$\stot= 62.17\pm 1.5$ mb.
The main purpose of the UA4/2 experiment was the measurement of the $\rho$
value, defined by
$\rho=\frac{\Re e f_n(t=0)}{\Im m f_n(t=0)}$%
where $f_n(t=0)$ is the forward {\em nuclear} scattering amplitude.
\subsection{Spinless Analysis Neglecting Magnetic
Scattering}\label{sec:spinless}

The experimenters parameterized the nuclear slope amplitude as
$%
f_n(|t|)=\frac{\sigma_{\rm tot}(\rho + i)}{4\sqrt{\pi}}e^{-b|t|/2},$%
\label{fn}
and measured the nuclear slope parameter as $b=15.5\pm 0.2$ GeV$^{-2}$.  They
fit $\pbar p$
elastic scattering data at $\sqrt s=541$ GeV over the  t-interval
$0.00075\le |t| \le 0.12$ GeV$^{2}$.
They used for the Coulomb amplitude the `spinless' Rutherford amplitude,
modified by a
Coulomb phase factor ${i\alpha \phi(t)}$, \ie
$f_c(t)= \frac{2\sqrt{\pi}\alpha}{|t|}G^2(t)e^{i\alpha \phi(t)},$%
where the phase\cite{cahnphase} $ \phi(t)$ is given by
\beq
\phi(t)=\mp \left\{\gamma +\ln \left(\frac{b|t|}{2}\right)
     +\ln\left(1+\frac{8}{b\Lambda^2}\right)+\left(\frac{4|t|}{\Lambda^2}
\right )
     \ln\left(\frac{4|t|}{\Lambda^2}\right)+\frac{2|t|}{\Lambda^2}\right\},
\label{cahn}
\eeq
where $\gamma=0.577\ldots$ is Euler's constant, $b$ is the slope parameter,
$\Lambda^2[=0.71$
GeV$^2$] appears in the dipole fit to the proton's electromagnetic form
factor,
$G(t)$. The upper
sign is for $pp$ and the lower sign for $\bar p p$.
Using these parameterizations,  the differential elastic scattering cross
section is
\begin{eqnarray}
\frac{d\sigma}{d|t|}=\left|f_c+f_n\right|^2%
=\frac{4\pi\alpha^2}{t^2}G^4(t)+\frac{\alpha\sigma_{\rm tot}}{|t|}(\rho +
\alpha \phi(t))G^2(t)e^{-b|t|/2}+\frac{\sigma_{\rm tot}^2(1+\rho^2)}{16\pi}e^{-
b|t|}.
\label{sumamp}
\end{eqnarray}
We now introduce the parameter $t_0$, defined as the absolute value of $t$
where the nuclear and
Coulomb amplitudes have the same magnitude, {\em i.e.,}
$
t_0=\frac{8\pi\alpha}{\sigma_{\rm tot}}=\frac{1}{14.00\sigma_{\rm tot}},
$
when $\sigma_{\rm tot}$ is in mb, and $t_0$ is in GeV$^2$. For $\sigma_{\rm
tot}= 62.17$ mb,
we find that $t_0=0.00115$ GeV$^2$.
We can now rewrite the differential cross section as
\beq
\frac{d\sigma}{d|t|}=\frac{\sigma_{\rm
tot}^2}{16\pi}\left\{G^4(t)\frac{t_0^2}{t^2}
+2\frac{t_0}{|t|}(\rho +\alpha\phi(t))G^2(t)e^{-b|t|/2}+(1+\rho^2)e^{-
b|t|}\right\},
\label{newdsdt}
\eeq
for $t_{\rm min}\le t \le t_{\rm max}$,
which was the form used by the UA4/2 group to analyze their experimental data.
They extracted
the value $\rho=0.135\pm 0.015$, with a statistical error of 0.007. We
emphasize that their
analysis, using \eq{newdsdt}, neglected the effects of the anomalous magnetic
moments of the
nucleons.
\subsection{`Spinless' Analysis, Taking into Account the  Magnetic Scattering}
For our small $|t|$ analysis, we approximate $G^4(t)$ as
$G^4(t)\approx 1-2a|t|,$
where $a\equiv \frac{4}{\Lambda^2}=5.6338$.
We further  write
$G^2(t)e^{-b|t|/2}\approx 1-(a+\frac{b}{2})|t|$.
However, if we take into account the anomalous magnetic moments, we see from
\eq{epsmallt2}
that we could have written
$f_c(t)
\approx 2\sqrt{\pi} \frac{\alpha}{|t|}G^2(t)\left\{
1+1.93 |t|\right\}e^{i\alpha\phi(t)}
,$%
 an energy-independent result. Literally, we have used the {\em `spinless'
ansatz} that the
Coulomb amplitude is given by the {\em square root} of the Coulomb cross
section.
Expanding the above equation in $|t|$, incorporating the new factor of $1+1.93
|t|$,
we can rewrite in a concise form the `correct' Coulomb amplitude as
\beq
f_c(t)
\approx 2\sqrt{\pi} \frac{\alpha}{|t|}G_{\rm eff}^2(t)e^{i\alpha\phi(t)}
,\label{fcnewexpand}
\eeq
where
\beq
G_{\rm eff}^2(t)\approx 1-a'|t|=1-(a-1.93)|t|=1-3.674|t|.
\eeq
Thus, $a'=3.674$, and we can now write a modified  form for the cross section
which mimics
\eq{newdsdt}
as
\beq
\frac{d\sigma'}{d|t|}=\frac{\sigma_{\rm tot}^2}{16\pi}\left\{G_{\rm
eff}^4(t)\frac{t_0^2}{t^2}
+2\frac{t_0}{|t|}(\rho ' +\alpha\phi(t))G_{\rm eff}^2(t)e^{-b|t|/2}+(1+\rho
'^2) e^{-b|t|}
\right\},\label{finaldsdt},
\eeq
for $t_{\rm min}\le |t|\le t_{\rm max}$,
by using $G_{\rm eff}(t)$ in place of $G(t)$, {\it i.e.,} by replacing $a$ by
$a'$ and
$\rho$ by $\rho'$.

In order to extract the new value of $\rho'$ from \eq{finaldsdt} without
having
to refit
directly the experimental data, we will require that the {\em integral} of
\eq{finaldsdt}
be equal
to the {\em integral} of \eq{newdsdt}, from $t_{\rm min}$ to $t_{\rm max}$.
This insures that
we fit the {\em measured events} with both formulae, allowing us to solve for
$\rho'$, given
the UA4/2 published
value of $\rho=0.135$. Thus, we require that
\beq
\int_{t_{\rm min}}^{t_{\rm max}}\frac{d\sigma'}{d|t|}=\int_{t_{\rm
min}}^{t_{\rm max}}
\frac{d\sigma}{d|t|}\label{equate}.
\eeq
To obtain an approximate analytical solution, we can set equal the
contributions of the terms
$(1+\rho '^2) e^{-b|t|}$ and $(1+\rho ^2) e^{-b|t|}$, and expand in $|t|$ to
first order. We
find the
approximate solution for $\rho'$, the {\em corrected} value of $\rho$, valid
when we set the
contribution of $\alpha\phi(t)=0$,
\begin{eqnarray}
\rho'&=&\frac{(a'-a)t_0+\rho\left(1-(a+b/2)\frac{(t_{\rm max}-t_{\rm min})}
{\ln(t_{\rm max}/t_{\rm min})}\right)}{1-(a'+b/2)\frac{(t_{\rm max}-t_{\rm
min})}
{\ln(t_{\rm max}/t_{\rm min})}}\nonumber \\
&\approx&\rho +(a'-a)\left( t_0+\rho \frac{t_{\rm max}-t_{\rm min}}
{\ln(t_{\rm max}/t_{\rm min})}\right)\qquad{\rm if }\quad t_0\ll\rho
\frac{t_{\rm max}-
t_{\rm min}}
{\ln(t_{\rm max}/t_{\rm min})}
\label{approxrho}
\end{eqnarray}
We will later see numerically
that neglecting the $\alpha \phi(t)$ contribution is a reasonable
approximation, since the
average value of $\alpha\phi(t)=-0.00036\approx 0$ over the t-range in
question. We find, using
the UA4/2 extracted values described above, that $\Delta \rho=\rho'-\rho=\rho'-
0.135=-0.0114$.
An {\em exact numerical solution} of \eq{equate}, where we expand to first
order in $|t|$ yields
the final answer $\Delta \rho_{\rm UA4}=-0.0105$. This shift in $\rho$ is {\em
comparable} to the quoted total
error of $\pm 0.015$ and is {\em larger} than the statistical error of $\pm
0.007$.  The
solution in completely insensitive to
the slope parameter $b$, as well as to $\alpha\phi(t)$.  It mainly depends on
the value of
$\rho$
that's measured, and critically on the $t$ interval that was measured.
We note that these conclusions are in contrast with the analysis of this
problem made by
N.~Buttimore\cite{buttimore}, which did not take into account the experimental
$t$ interval.

Since all high energy
$\rho$ values come from experiments that measure about the same range of $t$
and similar $b$
values, it is likely that {\em all} the $\rho$ values for $\bar p p$
scattering
need to be
{\em lowered}
by $\sim 0.005$---$0.010$ and that all $pp$ values need to be {\em raised} by
about
he same amount. For example,
E710 measured\cite{E710} $\rho_{\bar pp}=0.134\pm 0.069$, $\sigma_{\rm
tot}=72.2\pm 2.7$
mb and $b=16.72$ GeV$^{-2}$,
with $t_{\rm min}=0.00075$ GeV$^2$ and $t_{\rm max}=0.077$ GeV$^2$, at $\sqrt
s=1.8$ TeV.
In this case, the exact numerical solution of \eq{equate} gives
$\Delta\rho_{\rm E710}=-0.004$.
Thus, any global fit to the $\rho$ values of $\bar pp$ and $pp$ scattering
should take these {\em systematic shifts} in the $\rho$ values (at {\em all}
energies)
into account.
\subsection{A Nuclear Model Taking Account of Magnetic Scattering}
In the preceding Section, we emulated the `spinless' analysis by {\em
ignoring}
the
structure of the nuclear scattering, and {\em not distinguishing} between spin
flip
and non-spin flip of
the nuclear scattering, which we will now remedy. In this Section, we will
introduce a
toy nuclear model,
which mimics the known electromagnetic matrix element of \eq{Mkappa}, and
adds
{\em coherently}
with it. We thus write the total matrix element for coherent nuclear and
Coulomb scattering,
taking into account the Bethe phase as expressed by Cahn\cite{cahnphase}, as
\begin{eqnarray}
M'_t&=&\pm \frac{4\pi\alpha\exp^{i\delta \phi(t)}}{|t|}\bar u(p_f)\left[-
\kappa F_2\left(\frac{p_f+p_i}{2m}\right)_\mu +(F_1+\kappa F_2)
\gamma_\mu\right]u(p_i)\times\nonumber\\
&&\bar u(p'_f)\left[-\kappa F_2\left(\frac{p'_f+p'_i}{2m'}\right)^\mu +
(F_1+\kappa F'_2)
\gamma^\mu\right]u(p'_i)\quad + \nonumber\\
&&\frac{4\pi g}{m^2}\bar u(p_f)\left[-\kappa_N
H_2\left(\frac{p_f+p_i}{2m}\right)_\mu +
(H_1+\kappa_N H_2)
\gamma_\mu\right]u(p_i)\times\nonumber\\
&&\bar u(p'_f)\left[-\kappa_N H_2\left(\frac{p'_f+p'_i}{2m}\right)^\mu +
(H_1+\kappa_N H_2)
\gamma^\mu\right]u(p'_i),\label{MkappaplusNwithphi}
\end{eqnarray}
where now the upper sign is for $\bar pp$ and the lower sign is for $pp$
scattering, since we
introduced $|t|$ into \eq{MkappaplusNwithphi}. We have substituted
$e^2=4\pi\alpha$ and defined
the strong coupling analog of
$\alpha$ as $h^2=4\pi g$, where $h$ is the (complex) nuclear `charge',
$\kappa_N$ is the
nuclear `anomalous magnetic moment' and
$H_1(t)$ and $H_2(t)$ are the nuclear form factors. We have replaced the
electromagnetic
propagator $|t|$ by the nuclear propagator $m^2$. Later, we will use the
optical theorem to
fix the real and imaginary
portions of $h^2$ and will also introduce ``electric'' and ``magnetic''
nuclear
form factors
$G_{E_N}$ and $G_{M_N}$, analogous to \eq{GEGM}, with
\begin{eqnarray}
G_{E_N}(t) \equiv H_1(t)+\frac{\kappa_N t}{4m^2}H_2(t)\qquad{\rm and}\quad%
G_{M_N}(t)\equiv H_1(t)+\kappa_N H_2(t),\label{GEGMN}
\end{eqnarray}
and
\beq
\begin{array}{rclcl}
G_{E_N}(t)&=&G_N(t)&=&\displaystyle e^{bt/4} \\[7mm]
G_{M_N}(t)&=&(1+\kappa_N)G_N(t)&=&\displaystyle (1+\kappa_N)e^{bt/4}.
\end{array}\label{GEGMNparameter}
\eeq

The squaring of $M'_t$ in \eq{MkappaplusNwithphi} will give rise to three
terms,
\beq
|M_t|^2=|M_c|^2\pm 2|M_c|(\Re eM_N+\alpha\phi)+|M_N|^2, \label{Mtsq}
\eeq
since $e^{i\alpha \phi(t)}\approx 1+i\alpha\phi(t)$.  The first term of
\eq{Mtsq} corresponds
to pure Coulomb scattering, the last term to pure nuclear scattering and the
term
$2|M_c|(\Re eM_N+\alpha\phi)$ to the coherent interference cross section
between nuclear and
Coulomb amplitudes.  We note that the Dirac structure of {\em all three terms}
in \eq{Mtsq} is
the {\em same}, which greatly simplifies the evaluation.
We have already calculated the term $|M_c|^2$ and the substitution of $\alpha
\rightarrow g$,
$F_1\rightarrow H_1$,$F_2\rightarrow H_2$ and
$\frac{1}{|t|}\rightarrow\frac{1}{m^2}$ in the
Coulomb term gives us the nuclear term $|M_N|^2$. Thus we find by inspection
of
\eq{dsigmadtppsmallt} that the nuclear differential
cross section is given by
\begin{eqnarray}
\frac{d\sigma_N}{dt}
&=&4\pi \frac{g^2}{\beta_{\rm lab}^2m^4}\times\nonumber\\
&&\left\{\left(\frac{G_{E_N}^2(t)-\frac{t}{4m^2}G_{M_N}^2(t)}{1-
\frac{t}{4m^2}}\right)^2\left(
1+\frac{t}{2mE_{\rm lab}}\right)\right.\nonumber\\
&&+\,G_{M_N}^2(t)\frac{G_{E_N}^2(t)-\frac{t}{4m^2}G_{M_N}^2(t)}{1-
\frac{t}{4m^2}}
\frac{t}{2E_{\rm lab}^2}\nonumber\\
&&+\,\left.\left[G_{M_N}^4(t)
+\frac{1}{2}\left(\frac{G_{E_N}^2(t)-G_{M_N}^2(t)}{1-\frac{t}{4m^2}}\right)^2
\right]\frac{t^2}{8m^2E_{\rm lab}^2}\right\}
.\label{dsigmadtppN}
\end{eqnarray}
We now expand \eq{dsigmadtppN} for \Em{very small} $|t|$, using
\eq{GEGMNparameter}
and find that
\begin{eqnarray}
\frac{d\sigma_N}{dt}
&\approx &4\pi \frac{g^2}{\beta_{\rm lab}^2m^4}e^{bt}
\left\{1-\kappa_N(\kappa_N +2)\frac{t}{2m^2}+\frac{t}{2mE_{\rm lab}}
+(\kappa_N +1)^2\frac{t}{2E_{\rm lab}^2}\right\}
.\label{dsigmadtppsmalltN}
\end{eqnarray}
Using the optical theorem, we now rewrite
\beq
\left(\frac{d\sigma_N}{dt}\right)_{t=0}=\frac{\sigma_{\rm
tot}^2(1+\rho^2)}{16\pi}=
4\pi \frac{g^2}{\beta_{\rm lab}^2m^4}, \label{opttheorem1}
\eeq
since $G_N(0)\equiv 1$. Inspection of \eq{opttheorem1} yields the equivalent
statement that the (complex) value of the nuclear coupling is
$\frac{g}{\beta_{\rm lab}m^2}=(\rho +i)\frac{\sigma_{\rm
tot}}{8\pi}.$
 Using \eq{opttheorem1}, we can rewrite the nuclear  differential scattering
cross section of \eq{dsigmadtppsmalltN} as
\begin{eqnarray}
\frac{d\sigma_N}{dt}&=
&\frac{\sigma_{\rm tot}^2(1+\rho^2)}{16\pi}e^{bt}
\left\{1-\kappa_N(\kappa_N +2)\frac{t}{2m^2}+\frac{t}{2mE_{\rm lab}}
+(\kappa_N +1)^2\frac{t}{2E_{\rm lab}^2}\right\}
.\label{dsigmadtppsmalltN1}
\end{eqnarray}
The linear term in the brackets of
\eq{dsigmadtppsmalltN} or \eq{dsigmadtppsmalltN1} is clearly the spin flip
term
induced by the
nuclear ``anomalous magnetic
moment'' $\kappa_N$ and goes to zero in the forward direction, as must be true
for spin flip
amplitudes. At ultra-high energies, \eq{dsigmadtppsmalltN} goes over to
$\frac{d\sigma_N}{dt}=
\frac{\sigma_{\rm tot}^2(1+\rho^2)}{16\pi}e^{bt}
\left(1-\frac{\kappa_N(\kappa_N +2)}{2m^2}t\right)$

Again, with a lengthy, but straightforward calculation, we find that the
interference cross
section
$\frac{d\sigma_{\rm CN}}{dt}$ is given by
\begin{eqnarray}
\frac{d\sigma_{\rm CN}}{dt}&=&\mp \frac{\alpha}{\beta_{\rm lab}|t|}\sigma_{\rm
tot}( \rho +
\alpha\phi(t))\times\nonumber\\
&&\left\{
\left[
\left(\frac{G_E(t)-\frac{t}{4m^2}G_M(t)}{1-\frac{t}{4m^2}}\right)
\left(\frac{G_{E_N}(t)-\frac{t}{4m^2}G_{M_N}(t)}{1-
\frac{t}{4m^2}}\right)\right.\right.
\nonumber\\
&&\left.\left.\quad -\left(\frac{G_E(t)-G_M(t)}{1-\frac{t}{4m^2}}\right)
\left(\frac{G_{E_N}(t)-G_{M_N}(t)}{1-\frac{t}{4m^2}}\right)\frac{t}{4m^2}
\right]^2
\left(1+\frac{t}
{2mE_{\rm lab}}\right)
\right.\nonumber\\
&&\!\!\!\!\!\!+\,G_M(t)G_{M_N}(t)\frac{G_E(t)G_{E_N}(t)-
\frac{t}{4m^2}G_M(t)G_{M_N}(t)}
{1-\frac{t}{4m^2}}
\frac{t}{2E_{\rm lab}^2}\nonumber\\
&&\!\!\!\!\!\!+\,\left.
\left[\left(G_M(t)G_{M_N}(t)\right)^2
+\frac{1}{2}\left(\frac{G_E(t)G_{E_N}(t)-G_M(t)G_{M_N}(t)}{1-
\frac{t}{4m^2}}\right)^2
\right]\frac{t^2}{8m^2E_{\rm lab}^2}\right\}.\label{dsdtCN}
\end{eqnarray}
Taking the limit of \eq{dsdtCN} for small $t$, we find that
\begin{eqnarray}
\frac{d\sigma_{\rm CN}}{dt}&=&\mp \frac{\alpha G^2(t)}{\beta_{\rm
lab}|t|}\sigma_{\rm tot}
e^{bt/2}\left[ \vphantom{\frac{(\kappa +1)(\kappa_N +1)}{2E_{\rm lab}^2}}
\rho +\alpha\phi(t)\right]\times\nonumber\\
&&\left[1-\left(\frac{\kappa+\kappa_N+\kappa\kappa_N
}{2m^2}+\frac{1}{2mE_{\rm lab}}-\frac{(\kappa +1)(\kappa_N +1)}
{2E_{\rm lab}^2}\right)t\right].\label{dsdtCNlowt}
\end{eqnarray}
Introducing again the parameter
$t_0=\frac{8\pi\alpha}{\sigma_{\rm tot}}$, we write, in the small $t$ limit,
the  elastic
differential
scattering cross section $\frac{d\sigma}{d|t|}$ for {\em coherent} Coulomb and
nuclear
interactions as
\begin{eqnarray}
\frac{d\sigma}{d|t|}&=&\frac{\sigma_{\rm tot}^2}{16\pi}%
\left\{\frac{t_0^2}{t^2}\frac{1}{\beta_{\rm
lab}^2}G^4(t)\left[1+\left(\frac{\kappa(\kappa+2)}
{2m^2}-
\frac{1}{2mE_{\rm lab}}+\frac{(\kappa +1)^2}
{2E_{\rm lab}^2}\right)|t|\right]\right.\nonumber\\
&&\hphantom{\frac{\sigma_{\rm tot}^2}{16\pi}}\mp
2\frac{t_0}{|t|}\frac{1}{\beta_{\rm lab}}
G^2(t)
e^{-b|t|/2}\left[ \vphantom{\frac{(\kappa +1)(\kappa_N +1)}{2E_{\rm lab}^2}}
\rho +\alpha\phi(t)\right]\times\nonumber\\
&&\hphantom{\frac{\sigma_{\rm tot}^2}{16\pi}\mp}
\qquad\left[1+\left(\frac{\kappa+\kappa_N+\kappa\kappa_N
}{2m^2}-\frac{1}{2mE_{\rm lab}}+\frac{(\kappa +1)(\kappa_N +1)}
{2E_{\rm lab}^2}\right)|t|\right]\nonumber\\
&&\hphantom{\frac{\sigma_{\rm tot}^2}{16\pi}}
\left.+(1+\rho^2)e^{-b|t|}\left[1+\left(\frac{\kappa_N(\kappa_N+2)}{2m^2}-
\frac{1}{2mE_{\rm lab}}+\frac{(\kappa +1)^2}
{2E_{\rm lab}^2}\right)|t|\right]\right\}
\label{dsdtCNlowttot}
\end{eqnarray}
Finally, in the high energy limit, \eq{dsdtCNlowttot} simplifies to
\begin{eqnarray}
\frac{d\sigma}{d|t|}&=&\frac{\sigma_{\rm tot}^2}{16\pi}%
\left\{\frac{t_0^2}{t^2}G^4(t)\left[1+\frac{\kappa(\kappa+2)}{2m^2}
|t|\right]\right.\nonumber\\
&&\hphantom{\frac{\sigma_{\rm tot}^2}{16\pi}}
\mp 2G^2(t)\frac{t_0}{|t|}
\left( \vphantom{\frac{(\kappa +1)(\kappa_N +1)}{2E_{\rm lab}^2}}
\rho +\alpha\phi(t)\right)e^{-b|t|/2}
\left[1+\frac{\kappa+\kappa_N+\kappa\kappa_N
}{2m^2}|t|\right]\nonumber\\
&&\hphantom{\frac{\sigma_{\rm tot}^2}{16\pi}}
\left.+(1+\rho^2)e^{-b|t|}\left[1+\frac{\kappa_N(\kappa_N+2)}{2m^2}
|t|\right]\right\}
\label{dsdtCNlowttothigh}
\end{eqnarray}
In order to use \eq{dsdtCNlowttothigh}, we must expand the nuclear term for
small $t$
\begin{eqnarray}
\left(\frac{d\sigma}{d|t|}\right)_N&=&\frac{{\stot}^2}{16\pi}(1+\rho^2)e^{-
b|t|}
\left[1+\frac{\kappa_N(\kappa_N+2)}{2m^2}|t|\right]\nonumber\\
&\approx &\frac{{\stot}^2}{16\pi}(1+\rho^2)
\left[1-\left(b-\frac{\kappa_N(\kappa_N+2)}{2m^2}\right)|t|\right]
\nonumber\\
&\approx &\frac{{\stot}^2}{16\pi}(1+\rho^2)e^{-b'|t|},\qquad{\rm where
}\,\,\,b'=b-
\frac{\kappa_N(\kappa_N+2)}{2m^2}.
\end{eqnarray}
Effectively, it is $b'$ that is measured, {\em not} $b$. Thus we rewrite
\eq{dsdtCNlowttothigh}
as
\begin{eqnarray}
\frac{d\sigma}{d|t|}&\approx&\frac{\sigma_{\rm tot}^2}{16\pi}%
\left\{\frac{t_0^2}{t^2}G^4(t)\left[1+\frac{\kappa(\kappa+2)}{2m^2}
|t|\right]\right.\nonumber\\
&&\hphantom{\frac{\sigma_{\rm tot}^2}{16\pi}}
\mp 2G^2(t)\frac{t_0}{|t|}
\left( \vphantom{\frac{(\kappa +1)(\kappa_N +1)}{2E_{\rm lab}^2}}
\rho +\alpha\phi(t)\right)e^{-b'|t|/2}
\left[1+\frac{2\kappa+2\kappa\kappa_N-\kappa_N^2
}{4m^2}|t|\right]\nonumber\\
&&\hphantom{\frac{\sigma_{\rm tot}^2}{16\pi}}
\left.+(1+\rho^2)e^{-b'|t|}\right\}
\label{dsdtCNlowttothighnew}
\end{eqnarray}

We really have no deep knowledge of a value that is
appropriate for $\kappa_N$, let alone its sign.  Since the known polarization
at Fermilab
energies\cite{fermilab,leader} is small, of the order of several percent, a
magnitude
compatible with it being produced by Coulomb interactions, this suggests that
$|\kappa_N|\,  {}_\sim^<\, \kappa$. It is tempting, however,
to set $\kappa_N=\kappa$, since the conventional interpretation of the
anomalous magnetic
moment $\kappa$ is that it arises from the strong interactions.  If we equate
$\kappa$
and $\kappa_N$, we find the {\em same} interference term of  \eq{finaldsdt}
where we ``ignored'' spin, {\em i.e.,} $2G^2(t)\frac{t_0}{|t|}e^{-
b|t|/2}(1+1.93|t|)$ .
Thus, the numerical evaluation made earlier in Section \ref{sec:spinless} is
valid---namely,
for UA4/2, the $\rho$ value changes by $\approx -0.01$.

On the other hand, even if  we set $\kappa_N=0$,
we  {\em still} would have an interference term proportional to $(1+1.02|t|)$.

We see from \eq{approxrho} that
for $\kappa_N =0$ that the shift in $\rho$, which is proportional to $a'-a$,
is now reduced
by $1.02/1.93=0.53$. Thus, the $\rho$ value for $\kappa_N=0$ is shifted by
$\approx -0.006$.
We conclude that for reasonable values of the parameter $\kappa_N$, there is a
\Em{significant} shift in the $\rho$ value of UA4/2 due to the anomalous
magnetic moment of
the nucleon.  The
{\em theoretical uncertainty} in the $\rho$-value can be reduced if the
experimenters
break their data up into {\em two} distinct regions---Region 1 being the
interference
region from
$t_{\rm min}$ to $\approx 10\times t_0$ and Region 2 from $10\times t_0$ to
$t_{\rm max}$.
In the case of UA4/2, this would change the uncertainty in $\Delta \rho_{\rm
UA4}$ from
$\approx 0.006$ to $\approx 0.001$, if we found $\rho$ from Region 1 and
$\stot$ and $b$
from Region 2, even assuming that $\kappa_{N}=0$.

The other theoretical uncertainty is in the value of $\alpha \Phi(t)$.  The
original
derivation\cite{cahnphase} assumed that we had `spinless' scattering, that the
nuclear
amplitude was $e^{b't/2}$, where $b'$ was the measured slope, and that the
Coulomb
amplitude was $\frac{\alpha}{t}G^2(t)$.  We see that we really should be using
$\frac{\alpha}{t}G_{\rm eff}^2(t)$, or, using $\Lambda'^2\approx 1.0$ rather
than
$\Lambda^2= 0.71$, in \eq{cahn}.  Fortunately, this is a {\em very small}
change,
and increases
the $\pbar p$ $\rho$-value by $\approx 0.001$, thus contributing negligibly to
$\Delta \rho_{\rm UA4}$.

In conclusion, it seems sensible for experimenters to redo their data analysis
using
$\kappa_N=\kappa$, \ie using $G_{\rm eff}(t)$ and $\Lambda'^2$  rather than
$G(t)$ and
$\Lambda^2$,  in  the two distinct $t$-regions described above. This procedure
allows
the experimenter to control  the theoretical uncertainties, \ie our lack of
knowledge
of the nuclear amplitudes.
\section{Acknowledgements}
I would like to thank Dr. Nigel Buttimore for interesting discussions and Dr.
Robert N. Cahn for
sound advice and stimulating suggestions.

\end{document}